\documentclass[pre,
               twocolumn,
               showpacs,
               superscriptaddress
               ]{revtex4-1}

\usepackage{graphicx}
\usepackage{dcolumn}
\usepackage{bm}
\usepackage{amsmath}
\usepackage{amssymb}
\usepackage{color}

\begin{document}

\title{Inertial waves and mean velocity profiles in a rotating pipe \\
       and a circular annulus with axial flow}

\author{Yantao Yang}\email{yantao.yang@utwente.nl}
\affiliation{Physics of Fluids Group, Faculty of Science and Technology, MESA+ Research
             Institute, and J. M. Burgers Centre for Fluid Dynamics,
             University of Twente, PO Box 217, 7500 AE Enschede, The Netherlands.}
\affiliation{State Key Laboratory of Turbulence and Complex System, College of Engineering, \\
             Peking University, Beijing, 100871, China}

\author{Rodolfo Ostilla-M\'{o}nico}
\affiliation{Physics of Fluids Group, Faculty of Science and Technology, MESA+ Research
             Institute, and J. M. Burgers Centre for Fluid Dynamics,
             University of Twente, PO Box 217, 7500 AE Enschede, The Netherlands.}

\author{J.Z. Wu}
\affiliation{State Key Laboratory of Turbulence and Complex System, College of Engineering, \\
             Peking University, Beijing, 100871, China}

\author{P. Orlandi}
\affiliation{Dipartimento di Ingegneria Meccanica e Aerospaziale, Universit\`{a} La Sapienza, Roma, Italy}

\date{\today}

\begin{abstract}
In this paper we solve the inviscid inertial wave solutions in a circular pipe or annulus rotating constantly about its axis with moderate angular speed. The solutions are constructed by the so-called helical wave functions. We reveal that the mean velocity profiles must satisfy certain conditions to accommodate the inertial waves at the bulk region away from boundary. These conditions require the axial and azimuthal components of the mean velocity take the shapes of the zeroth and first order Bessel functions of the first kind, respectively. The theory is then verified by data obtained from direct numerical simulations for both rotating pipe and circular annulus, and excellent agreement is found between theory and numerical results. Large scale vortex clusters are found in the bulk region where the mean velocity profiles match the theoretical predictions. The success of the theory in rotating pipe, circular annulus, and streamwise rotating channel suggests that such inertial waves are quite common in wall bounded flow with background rotation.    
\end{abstract}


\maketitle

\section{Introduction}

Flows with background rotation are ubiquitous in nature and engineering environments. For example, the large scale motions in atmosphere and ocean are inevitably under the influences of Earth rotation, and rotary machine is very common in modern engineering. A unique characteristic of rotating fluids is that it carries inertial waves (IWs), such as the planar IWs in infinite domains~\cite{Greenspan:1969}. In confined domains, the boundary introduces new constrains for the existence of IWs and therefore modify their properties. IWs may be reflected by the wall and change wavenumbers after the reflection~\cite{Phillips:1963}. The IWs have been detected in many experiments, such as rotating grid turbulence~\cite{Bewley_etal:iwave:2007,Lamriben_etal:PoF:2011,Boisson_etal:PoF:2012} and spherical Couette flow~\cite{Kellyetal:PRE,Rieutord_etal:PRE:2012}, to name a few. For rotating grid turbulence, the IWs have large scales and resonate at characteristic frequencies of the tank~\cite{Bewley_etal:iwave:2007}, which quantitively match the theoretical predictions given by Batchelor~\cite{Batchelor:1967} and Maas~\cite{Maas:iwave}. In the recent experiment of a rotating water-filled cylinder~\cite{Messio_etal:iw:2008}, the vortical structures associated with the IWs were illustrated using flow fields obtained by particle image velocimetry.

IWs have also been investigated in many theoretical works, such as the asymptotic properties of inertial modes in a rotating spherical shell~\cite{Rieutordetal:JFM:2001} and the elliptic instability of a Rankin vortex with axial mean flow~\cite{Lacaze_etal:PoF:2005}. The inertial mode solutions have been constructed semi-analytically for a rectangular parallelepiped~\cite{Maas:iwave,Nurijanyan_etal:PoF:2013}. When a background current is added to the system, IWs may show different behaviors. For the simplest case, Greenspan discussed the IWs in a uniform current with constant velocity~\cite{Greenspan:1969}. The uniform current induces a Doppler effect, and thus changes the frequency and the phase velocity. The group velocity, which indicates the direction of energy transfer, may even be against the current direction provided the wave number meets certain conditions~\cite{Greenspan:1969}. 

In our recent paper~\cite{Yangetal:JFM:2010}, we studied the IWs in a channel rotating around a streamwise axis. The inviscid IW solutions were constructed by the so-called helical wave functions. We showed that IW solutions may still exist when there exists a mean flow. And IW can be strong enough to modify the profile of the mean streamwise velocity and induce a secondary mean flow in the spanwise direction. Using the concept of IW, we successfully explained the shapes of the mean velocity profiles. The numerical results support our theoretical predictions and the existence of IWs in streamwise rotating channel flow. The flow visualisation reveals large-scale vortical structures for moderate rotating rate, which are associated with the IWs~\cite{Yangetal:JFM:2010}.

Here we will investigate the IWs in a rotating circular pipe or annulus with an axial mean flow. The methodology used here is similar to that for channel flow~\cite{Yangetal:JFM:2010}. We mainly focus on the existence of the IWs and the associated mean velocity profiles. The paper is organised as follows. In Sec.~\ref{sec:theory} we will provide the theoretical formulation of the IW solutions in cylindrical domain. In Sec.~\ref{sec:results} we will validate the theory by the numerical results of rotating pipe and circular annulus. And finally the conclusions are given in Sec.~\ref{sec:conclusion}.
 
\section{Theoretical formulation}\label{sec:theory}

\subsection{Inviscid inertial wave solution without axial mean flow}

We now construct the inertial wave solution from the helical wave functions. Suppose the pipe or annulus rotates around its axis with a constant angular velocity $\bm{\Omega} = \Omega \bm{e}_z$. In the rotating frame of reference, the incompressible Navier-Stokes equation may be written as
\begin{equation}\label{eq:NS}
    \frac{\partial \bm{u}}{\partial t} + \bm{\omega}\times\bm{u} + 2\bm{\Omega}\times\bm{u} = -\nabla P_\text{m} + \nu \nabla^2 \bm{u},
\end{equation}
in which $\bm{u}$ is velocity, $\bm{\omega} = \nabla\times\bm{u}$ is vorticity, and $\nu$ is the kinematic viscosity, respectively. The modified pressure $P_\text{m} = p/\rho + |\bm{u}|^2 / 2 + \Pi_c$ with $p$ being pressure, $\rho$ the density, and $\Pi_c = [(\mathbf{\Omega}\cdot\bm{r})^2 - \Omega^2 r^2] / 2$ the centrifugal potential. The nonlinear term $\bm{l}=\bm{\omega}\times\bm{u}$ is also known as the Lamb vector. By taking the curl of \eqref{eq:NS}, one gets the vorticity equation
\begin{equation}\label{eq:vort}
    \frac{\partial \bm{\omega}}{\partial t} + \nabla\times\bm{l} = 2\bm{\Omega}\cdot\nabla\bm{u} + \nu\nabla^2\bm{\omega}.
\end{equation}
Two important non-dimensional parameters are the Rotation number $Ro = 2\Omega R / U$ and the Reynolds number $Re = UR/\nu$. Here $U$ is some characteristic velocity. The Rotation number measures the ratio of the Coriolis force to the inertial term. The larger $Ro$ is, the faster the domain rotates.

The helical wave functions have long been used to construct the inviscid IW solution in the infinite domain, c.f. Refs.~\cite{Greenspan:1969} and \cite{Waleffe:1993}. Here we apply them to the cylindrical domain. A brief description of the helical wave functions in cylindrical domain is given in Appendix~\ref{app:hwf}. Let us first consider the inviscid inertial oscillation with no background flow, i.e. the wave-form solution of \eqref{eq:NS} and \eqref{eq:vort} with $\nu = 0$. Assume the inertial oscillation has form
\begin{equation}\label{eq:iwsolution}
  \left\{
  \begin{array}{rcl}
    \displaystyle \bm{u} & = & \bm{\varphi}_s\exp[i(m\theta + k_z z-f_s t)], \\[0.2cm]
    \displaystyle \bm{\omega} & = & sk \bm{\varphi}_s\exp[i(m\theta + k_z z-f_s t)],
  \end{array}
  \right.
\end{equation}
where $f_s$ is the frequency to be determined, $k^2=k_r^2+k_z^2$ with $k_r$ and $k_z$ being the wave numbers in the radial and axial directions, and $s=\pm 1$ is the polarity index of the function, respectively. Note that such a choice of $(\bm{u}, \bm{\omega})$ ensures $\bm{l} \equiv \bm{0}$. So that the inviscid form of \eqref{eq:vort} reduces to 
\begin{equation}\label{eq:vort-iw}
    \frac{\partial \bm{\omega}}{\partial t} = 2\bm{\Omega}\cdot\nabla\bm{u}.
\end{equation}
By substituting \eqref{eq:iwsolution} into \eqref{eq:vort-iw} one obtains the dispersion relation as
\begin{equation}\label{eq:dispersion}
  f_s = - \frac{2 s k_z \Omega}{k}.
\end{equation}
The phase velocity has azimuthal and axial components, which are
\begin{subequations}\label{eq:cp}
\begin{eqnarray}
  \displaystyle  c_{p\theta} (r) &=& -\frac{2sk_z\Omega}{k m}r, \label{eq:cpt}  \\[0.2cm]
  \displaystyle  c_{pz} &=& -\frac{2s\Omega}{k}. \label{eq:cpz}
\end{eqnarray}
\end{subequations}
$c_{p\theta}$ is proportional to $r$ so that the phase keep constant in every meridian plane as the wave transfers in the azimuthal direction. The group velocity is
\begin{equation}\label{eq:cg}
   c_{gz} = \frac{\partial f_s}{\partial k_z} = -2s\,\Omega \frac{k_r^2}{k^3}.
\end{equation}

The IWs \eqref{eq:iwsolution} share several common properties with the planar ones in the infinite domain. For instance, the dispersion relation indicates that the frequency depends on the angle between the wavenumber vector and the rotating axis, but is independent with the magnitude of the wavenumber vector. Equation~\eqref{eq:cpz} implies that long waves with smaller $k$ travel faster in the axial direction than short waves with larger $k$. One difference is that IWs propagate energy along the pipe axis as they travel in $(\theta,z)$-plane since the group velocity only has an axial component,  while for planar IWs the group velocity is always perpendicular to the wavenumber vector~\cite{Greenspan:1969}.

\subsection{Effects of axial mean flow}

Now consider the flow driven by an external constant mean pressure gradient $\partial \overline{P} / \partial z$ along the axis. Hereafter an overline stands for averaging in $z$- and $\theta$-directions and over time. Based on the existing experiments and numerical simulations,  it is reasonable to assume that the mean flow has the velocity and vorticity profiles
\begin{equation}\label{eq:mflow}
\begin{array}{rcl}
 \displaystyle \overline{\bm{u}}(r) &=&  \left(\, 0,\ \overline{u}_\theta(r),\ \overline{u}_z(r) \,\right), \\[0.1cm]
 \displaystyle \overline{\bm{\omega}}(r) &=& 
         \left(\, 0,\ \overline{\omega}_\theta(r),\ \overline{\omega}_z(r) \,\right) \\[0.1cm]
  &=& \left(\,0,\ -\partial_r \overline{u}_z,\ \partial_r(r\overline{u}_\theta) / r \,\right), \\
\end{array}
\end{equation}
which depend solely on $r$ and have zero radial components. $\partial_r$ represents the derivative with respect to $r$. 

We denote the fluctuating velocity and vorticity by $\bm{u}'$ and $\bm{\omega}'$, respectively. Following the methodology of Refs.~\onlinecite{Waleffe:1993} and \onlinecite{Davidson:2004}, we seek the travelling wave solution of the inviscid fluctuating vorticity equation
\begin{eqnarray}
    && \hspace{-1cm} \frac{\partial \bm{\omega}'}{\partial t} + \nabla\times\bm{L} 
       + \bm{u}'\cdot\nabla\,\overline{\bm{\omega}} - \bm{\omega}'\cdot\nabla\,\overline{\bm{u}} \nonumber \\
    && \hspace{1cm} = 2\bm{\Omega}\cdot\nabla\,\bm{u}' 
    + \overline{\bm{\omega}}\cdot\nabla\,\bm{u}' - \overline{\bm{u}}\cdot\nabla\,\bm{\omega}',\label{eq:fluc-om}
\end{eqnarray}
where $\bm{L} = \bm{\omega}'\times\bm{u}' - \overline{\bm{\omega}'\times\bm{u}'}$. We have made the assumption that $Re$ is large so that the viscous effect may be neglected at the bulk region away from the wall boundary.

At moderate rotation rate with $Ro \sim 1$, in \eqref{eq:fluc-om} the terms containing mean quantities have comparable order of magnitude with the Coriolis term, thus they are not negligible. However, one can still seek the IW solutions of the equation. Assume that the formal wave solution \eqref{eq:iwsolution} exists, then for such velocity and vorticity fields one has $\bm{\omega}'=sk\,\bm{u}'$ and $\bm{L}\equiv\bm{0}$. Eq.~\eqref{eq:fluc-om} can be rewritten as
\begin{equation}
     \hspace{-1cm} \frac{\partial \bm{\omega}'}{\partial t} 
        + 2\bm{u}'\cdot\nabla\, \bm{\Omega}^\text{ad}  
      = 2(\bm{\Omega}+\bm{\Omega}^\text{ad})\cdot\nabla\,\bm{u}', \label{eq:fluc-omb}
\end{equation}
with $2\bm{\Omega}^\text{ad} = \overline{\bm{\omega}} - sk \,\overline{\bm{u}}$ being an added system rotation introduced by the mean flow. Note that $\bm{\Omega}^\text{ad}$ only has azimuthal and axial components and is a function of $r$. The component form of \eqref{eq:fluc-omb} reads
\begin{subequations}\label{eq:vort-iw-mf}
\begin{eqnarray}
 \hspace{-0.6cm} \frac{\partial \omega'_r}{\partial t} &=& 
         2 (\bm{\Omega}+\bm{\Omega}^\text{ad})\cdot\nabla u'_r, \\
 \hspace{-0.6cm} \frac{\partial \omega'_\theta}{\partial t} 
         + \underline{2\bm{u}'\cdot\nabla\Omega^\text{ad}_\theta} 
      &=& 2(\bm{\Omega}+\bm{\Omega}^\text{ad})\cdot\nabla u'_\theta 
         + \frac{\Omega^\text{ad}_\theta u'_r}{r}, \\
 \hspace{-0.6cm} \frac{\partial \omega'_z}{\partial t} 
         + \underline{2\bm{u}'\cdot\nabla\Omega^\text{ad}_z} 
      &=& 2(\bm{\Omega}+\bm{\Omega}^\text{ad})\cdot\nabla u'_z.
\end{eqnarray}
\end{subequations}
If the two terms with underline vanish, i.e. the mean flow satisfying the IW condition
\begin{equation}\label{eq:IWC1}
   \nabla\Omega^\text{ad}_\theta = \nabla\Omega^\text{ad}_z = \bm{0},
\end{equation} 
Eq.~\eqref{eq:vort-iw-mf} reduces to the same form as \eqref{eq:vort-iw} with $\bm{\Omega}$ being replaced by an effective rotation rate $\bm{\Omega}^\text{ef} = \bm{\Omega} + \bm{\Omega}^\text{ad}$. Meanwhile, \eqref{eq:IWC1} implies immediately that $\bm{\Omega}^\text{ad}$ is independent of $r$, then $\bm{\Omega}^\text{ef}$ is also constant. For the IW with axial mean flow, the dispersion relation and the expression for phase and group velocity have the same forms as \eqref{eq:dispersion}-\eqref{eq:cg}, only with $\Omega$ being replaced by $\Omega^\text{ef}$.

Eq.~\eqref{eq:IWC1} also determines the profiles of the mean velocity. Recall that both $\Omega^\text{ad}_\theta$ and $\Omega^\text{ad}_z$ only depend on $r$, thus by \eqref{eq:mflow} the explicit form of \eqref{eq:IWC1} reads
\begin{subequations}\label{eq:IWC2}
\begin{eqnarray}
  \displaystyle \partial_r \overline{\omega}_\theta &=& sK \partial_r \overline{u}_\theta, \label{eq:IWC2a} \\[0.1cm]
  \displaystyle \partial_r \overline{\omega}_z &=& sK \partial_r \overline{u}_z. \label{eq:IWC2b}
\end{eqnarray}
\end{subequations}
Here we use $K$ to denote the wavenumber which satisfies the condition \eqref{eq:IWC1}. For the flow in pipe or between two concentric cylinders, it is reasonable to assume that $\overline{u}_z$ has a maximum, which we denote as $u_{z0}$ at radial location $r_0$. Then by \eqref{eq:mflow} one has $\overline{\omega}_\theta(r_0)=0$. Suppose the azimuthal velocity at $r_0$ is $u_{\theta0}$, then one may further choose a rotating frame of reference so that $\overline{u}_\theta(r_0)=0$, which is easily achieved by setting $\Omega^*=\Omega-u_{\theta0}/r_0$. In this new frame of reference, \eqref{eq:IWC2a} is equivalent to $\overline{\omega}_\theta=sK\overline{u}_\theta$. And by combining \eqref{eq:mflow} and \eqref{eq:IWC2} one has
\begin{equation}\label{}
    \frac{\partial^2 \overline{u}_\theta}{\partial r^2} + \frac{1}{r} \frac{\partial \overline{u}_\theta}{\partial r} + \left( K^2-\frac{1}{r^2} \right) \overline{u}_\theta = 0,
\end{equation}
which gives
\begin{equation}\label{eq:uth-iw}
    \overline{u}_\theta = sA J_1(K\tilde{r}),
\end{equation}
where $\tilde{r}=r-r_0$ and $J_1$ is the first order Bessel function of the first kind. $A>0$ is the amplitude of the IWs. We notice that \eqref{eq:uth-iw} has the same form as the Kelvin mode of zero wavenumber in the axial direction, e.g. see~\cite{Batchelor:1967}. The profile of the streamwise mean velocity is then readily derived from \eqref{eq:IWC2} as
\begin{equation}\label{eq:uz-iw}
    \overline{u}_z = A J_0(K\tilde{r}) + u_{z0} - A,
\end{equation}
where $J_0$ is the zeroth order Bessel function of the first kind. Finally, by transferring back to the original rotating frame of reference with angular speed $\Omega$, the mean velocity profiles read
\begin{subequations}\label{eq:iwo}
\begin{eqnarray}
    \overline{u}_\theta &=& sA J_1(K\tilde{r}) + u_{\theta0} r / r_0, \label{eq:iwo-a} \\
    \overline{u}_z &=& A J_0(K\tilde{r}) + u_{z0} - A, \label{eq:iwo-b}
\end{eqnarray}
\end{subequations}
which will be tested against numerical results in the following section.

\section{Application to numerical results}\label{sec:results}

\subsection{Rotating pipe}

To test the above theoretical predictions for rotating pipe flow, we employ the data from the direct numerical simulation (DNS) done by Orlandi's group~\cite{Orlandi:pipe:1997,Orlandi:1997,Orlandi:pipe:2000}. Let $R$, $\nu$, $U_p$, and $\Omega$ be the pipe radius, the kinematic viscosity, the streamwise mean velocity at centerline, and the background rotation, respectively. The Reynolds number is $Re = U_p R/\nu$, and the rotation number is $Ro = 2\Omega R/U_p$. For details of the numerical methods please see Refs.~\onlinecite{Orlandi:pipe:1997,Orlandi:1997,Orlandi:pipe:2000}. Here we shall use two sets of simulations with different Reynolds numbers, and for each $Re$ we consider several different rotation rates, see Tab.~\ref{tab:pipe}. The corresponding grid number $N_r$ in the radial direction is also listed.
\begin{table}
\caption{Summary of the flow parameters for rotating pipe and corresponding $K$, $A$, and $u_{z0}$. $Re$ is the Reynolds number, $Ro$ is the rotation number, and $N_r$ is the radial grid number, respectively. For the case name in the first column, the letter indicates low (L) or high (H) $Re$, and the same digit indicates the same $Ro$.}\label{tab:pipe}
 \begin{ruledtabular}
  \renewcommand{\arraystretch}{1.25}
  \begin{tabular}{ccccccc}
    Case & $Re$ & $Ro$ & $N_r$ & $u_{z0}$ & $K$ & $A$  \\
    \hline 
    L1  & 4900  & 0.5 & 48  & 0.716 & 3.91 & 0.095    \\
    L2  & 4900  & 1.0 & 48  & 0.753 & 3.60 & 0.168    \\
    L3  & 4900  & 2.0 & 96  & 0.810 & 3.67 & 0.265    \\
    L4  & 4900  & 5.0 & 96  & 0.903 & 3.82 & 0.402    \\[0.1cm]
    H2  & 10000 & 1.0 & 192 & 0.752 & 3.68 & 0.196    \\
    H3  & 10000 & 2.0 & 192 & 0.802 & 3.56 & 0.297    \\
    H4  & 10000 & 5.0 & 192 & 0.910 & 3.95 & 0.426    \\
  \end{tabular}
 \end{ruledtabular}
\end{table}

\begin{figure}
  \centering
  \includegraphics[width=6cm]{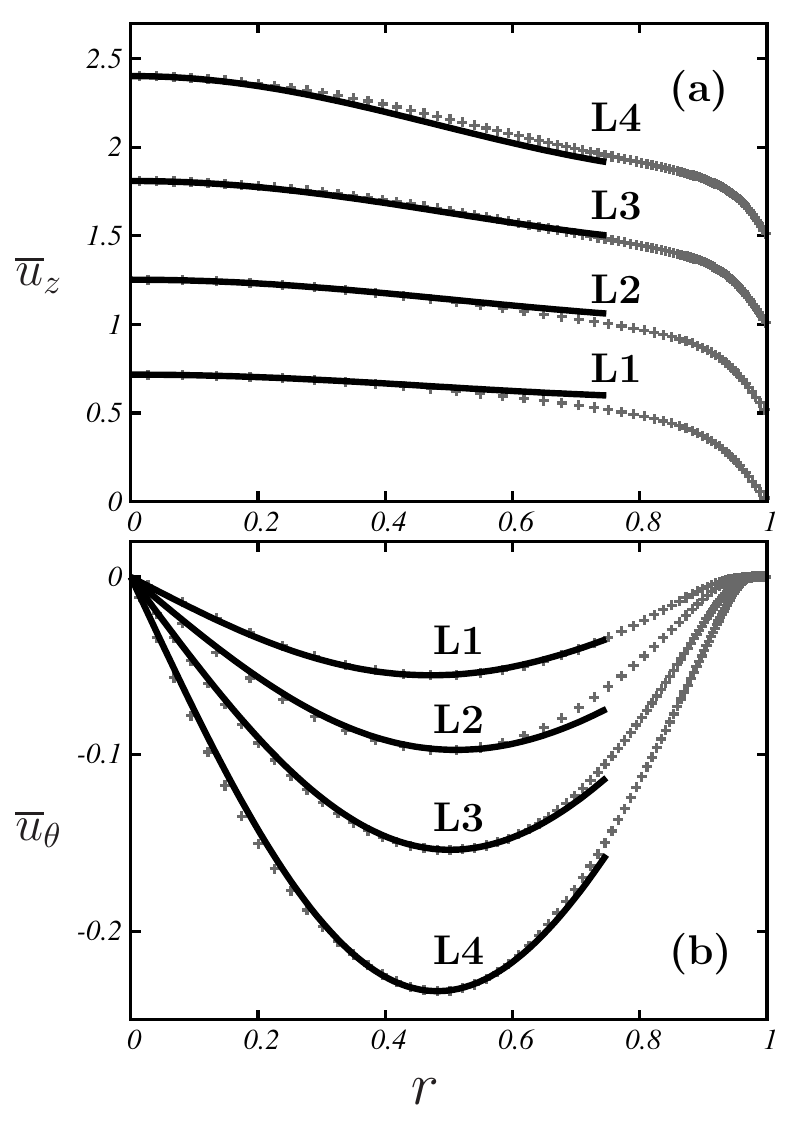}\\
  \caption{Velocity profiles of DNS results and the predictions of IW theory in the bulk region for low Reynolds-number cases. (a) $\overline{u}_z$ (symbols) compared to eq.~\eqref{eq:uz-iw} (lines). For clarity the curves are shifted upward from the previous one by $0.5$. (b) $\overline{u}_\theta$ (symbols) compared to eq.~\eqref{eq:uth-iw} (lines).}\label{fig:LowRe}
\end{figure}

\begin{figure}
  \centering
  \includegraphics[width=6cm]{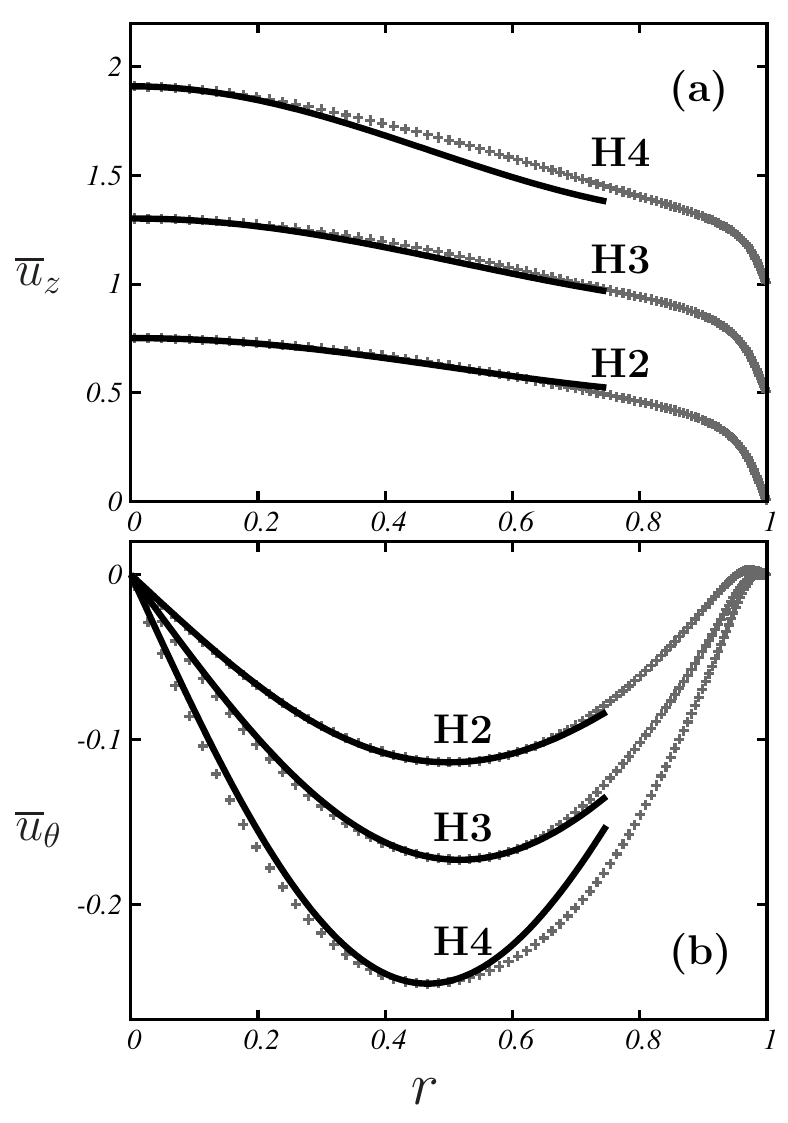}\\
  \caption{The same curves as in Fig.~\ref{fig:LowRe} for high Reynolds-number cases.}\label{fig:HighRe}
\end{figure}

The mean velocity profiles obtained by DNS are plotted in Figs.~\ref{fig:LowRe} and \ref{fig:HighRe} for low and high $Re$'s, respectively. A secondary mean flow develops in the azimuthal direction. We now validate the theory with these mean velocity profiles. First, $\overline{u}_z$ reaches maximum at the centreline, where $\overline{u}_\theta$ is zero. Thus $r_0=0$ and $\Omega^*=\Omega$. Then one can directly apply eqs.~\eqref{eq:uth-iw} and \eqref{eq:uz-iw}. $u_{z0}$ can be set to the mean axial velocity at the centreline. $\overline{u}_\theta$ is negative in the bulk region and reaches minimum around $r \approx 0.5$. Thus $s=-1$ and the IWs have negative polarity. The amplitude $A$ and wave number $K$ can be determined by the magnitude and location of the $\overline{u}_\theta$-minimum. Those parameters are also listed in Tab.~\ref{tab:pipe}. The theoretical curves so obtained are compared with the DNS results in the same figures for $\overline{u}_\theta$ and $\overline{u}_z$.

Figs.~\ref{fig:LowRe} and \ref{fig:HighRe} indicate that \eqref{eq:uth-iw} and \eqref{eq:uz-iw} are quite accurate in the bulk region for two small rotation rates. As $Ro$ increases, e.g. for $Ro = 2$ and $5$, the discrepancy between the theory and the simulations increases. It is well known that for very large $Ro$ the Coriolis term in \eqref{eq:fluc-om} dominates the momentum balance and the equation may be reduced to, by neglecting the viscous term in the bulk region,
\begin{equation*}
    \frac{\partial \bm{\omega}'}{\partial t} = 2\bm{\Omega}\cdot\nabla\bm{u}'.
\end{equation*}
Then \eqref{eq:iwsolution} with any wave number $k$ is an IW solution. No dominant wave number exists and the previous argument fails. When $Ro$ is around unity, however, the linear term with mean quantities cannot be neglected. The IW solutions can only survive at certain wave number $K$, and the mean flow evolves into certain profiles and satisfies the IW condition \eqref{eq:IWC1}. 

Another fact which can be observed from these two figures is that for a fixed $Ro$ the theory becomes less accurate as $Re$ increases, especially for high rotation number. A conclusive explanation about this needs more simulations with different parameters. One possible reason is that for higher $Re$ the viscous effect becomes weaker in the bulk region. Therefore, more inertial modes can be excited and the dominant wavenumber may vanish sooner as $Ro$ increases from $1$. 

The agreement between the DNS results and the theory strongly suggests that IW exists for $Ro$ around $1$. It should be pointed out that $K$ for different rotation numbers are very close to each other. As explained in Ref.~\onlinecite{Batchelor:1967}, the wave number of the inertial oscillation in a confined container is determined by the global properties, e.g. the geometry of the container. For the axial rotating pipe the wave number $K$ should be related to the diameter of the pipe, which is the same for all cases. Therefore, the value of $K$ should not vary too much for different $Ro$'s, as listed in Tab.~\ref{tab:pipe}. Recall that the secondary flow in the azimuthal direction is mainly induced by the inertial waves, a similar value of $K$ may also explain why the radial location where $u_\theta$ reaches its minimum is almost the same for all cases with $Ro$ near unit. Moreover, near the pipe boundary the viscous effects becomes dominant and the theory does not apply anymore. Thus as $r$ increases from the location of $u_\theta$-minimum towards pipe boundary, the discrepancy between the theory and simulations grows.

\subsection{Rotating circular annulus}

We now turn to the rotating circular annulus with an axial flow. Denote the radial gap of the annulus by $h$ and the axial bulk velocity by $U_b$, respectively. The annulus rotates around its axis with an constant angular speed $\Omega$. The Reynolds number and Rotation number are then defined as $Re = U_b h / \nu$ and $Ro = 2 \Omega h / U_b$. The flow is driven by an axial mean pressure gradient. The numerical solver is the same as those in Ref~\cite{Ostilla_etal:JFM:2013}. To save computation cost we only simulate one third of the whole annulus in the azimuthal direction, and periodic conditions are applied to both the axial and azimuthal directions. We fixed the Reynolds number at $Re=5000$ and simulated three different Rotation numbers, which are $Ro=0.5$, $1.0$, and $2.0$, respectively.
\begin{table}
\caption{The parameters for the flow in a circular annulus. For all cases $Re=5000$ and the radial grid number is $256$. $Ro$ is the rotation number, $r_0$ is the radial location of $\overline{u}_z$-maximum, $u_{z0}$ and $u_{\theta0}$ are the mean axial and azimuthal velocity at $r_0$, $(K_i, A_i)$ and $(K_o,A_o)$ are the wavenumber and amplitude for inner and outer IWs, respectively.}\label{tab:cyd}
\begin{ruledtabular}
\renewcommand{\arraystretch}{1.25}
\begin{tabular}{ccccccccc}
 Case & $Ro$ & $r_0$ & $u_{\theta0}$ & $u_{z0}$ & $K_i$  & $A_i$ & $K_o$ & $A_o$ \\ \hline 
 C1   & 0.5  & 1.434 & -0.0495 & 0.59 & 12.51 & 0.0343 & 10.92 & 0.0395   \\
 C2   & 1.0  & 1.439 & -0.0368 & 1.15 & 13.84 & 0.0396 & 11.64 & 0.0460   \\
 C3   & 2.0  & 1.439 & -0.0917 & 2.27 & 27.22 & 0.0117 & 12.92 & 0.0363   \\
\end{tabular}
\end{ruledtabular}
\end{table}

\begin{figure}
  \centering
  \includegraphics[width=6cm]{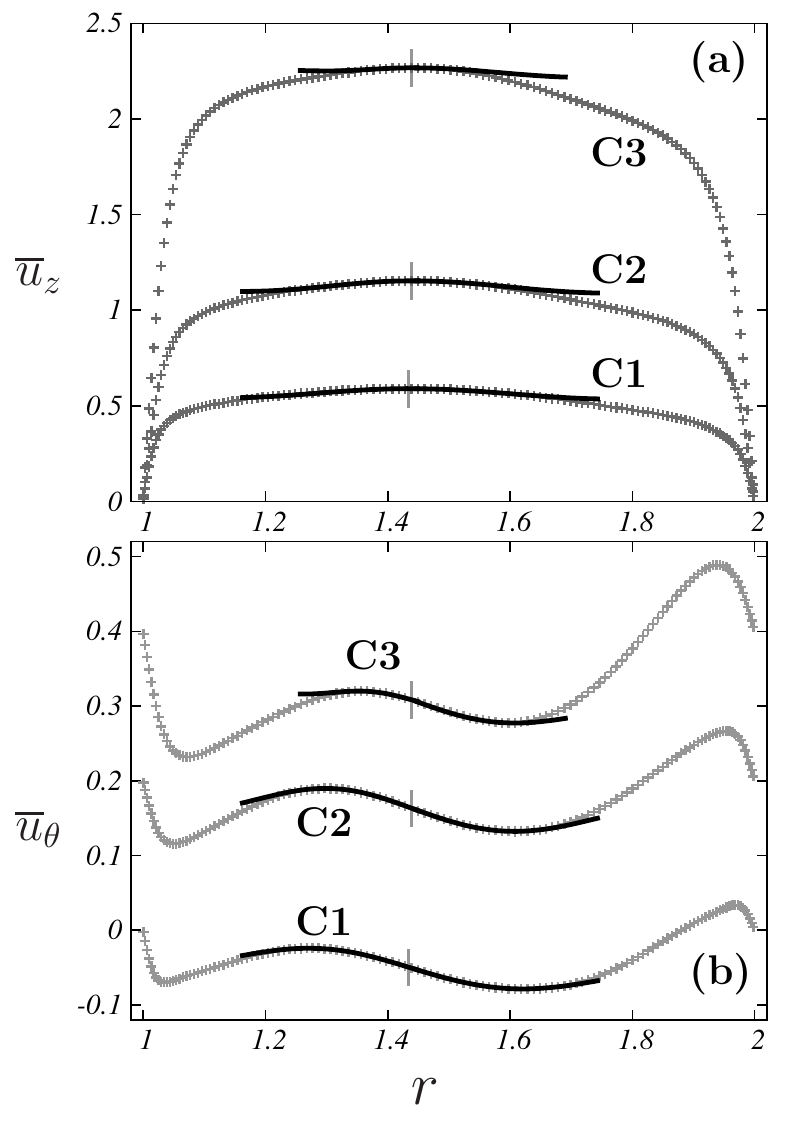}\\
  \caption{(Color online) The mean velocity profiles in rotating circular annulus. The symbols are numerical results and the curves are given by eq.~\eqref{eq:iwo}. The short vertical lines mark the locations $r_0$ of each case. The curves on the left of $r_0$ are computed by $(K_i, A_i)$ and on the right of $r_0$ by $(K_o, A_o)$, respectively. (a) profiles of mean axial velocity, and (b) profiles of mean azimuthal velocity. In (b) for clarity each curve is shifted upward by $0.2$ from the previous one.}\label{fig:RTC}
\end{figure}

In Fig.~\ref{fig:RTC} we plot the mean velocity profiles obtained by the numerical simulations. For all three cases, the mean axial velocity $\overline{u}_z$ reaches a maximum at similar location, $r_0\approx1.44$. $\overline{u}_{\theta}(r_0)$ is not equal to zero but a negative value. All those quantities are given in Tab.~\ref{tab:cyd}. In the bulk region $\overline{u}_\theta$ has an inner maximum at $r<r_0$ and an outer minimum at $r>r_0$ respectively. Near each boundary there is another peak which is in the opposite direction than the adjacent bulk peak. The overall shape of $\overline{u}_\theta$ is very similar to that in the streamwise rotating channel flow~\cite{Yangetal:JFM:2010}.    

To apply the theory to the current flow, we follow the same procedure as for pipe flow. However, extra care needs to be taken because now $\overline{u}_\theta(r_0)$ is not zero. Furthermore, the distance between the inner maximum of $\overline{u}_\theta$ and $r_0$ is not the same as that between the outer minimum and $r_0$. This suggests that the IW wavenumbers are different for $r<r_0$ and $r>r_0$. Thus by the inner maximum and the outer minimum we fix two sets of wavenumber and amplitude, i.e. $(K_i, A_i)$ and $(K_o,A_o)$, which are listed in Tab.~\ref{tab:cyd}. With these parameters we compute the theoretical predictions of the mean velocity profiles in the bulk region and compare to the numerical results, which are shown in Fig.~\ref{fig:RTC}. The agreement between the theory and the simulations are extremely well around $r_0$. Again, this strongly implies that IWs exist in such flow. 

\begin{figure}
  \centering
  \includegraphics[width=8cm]{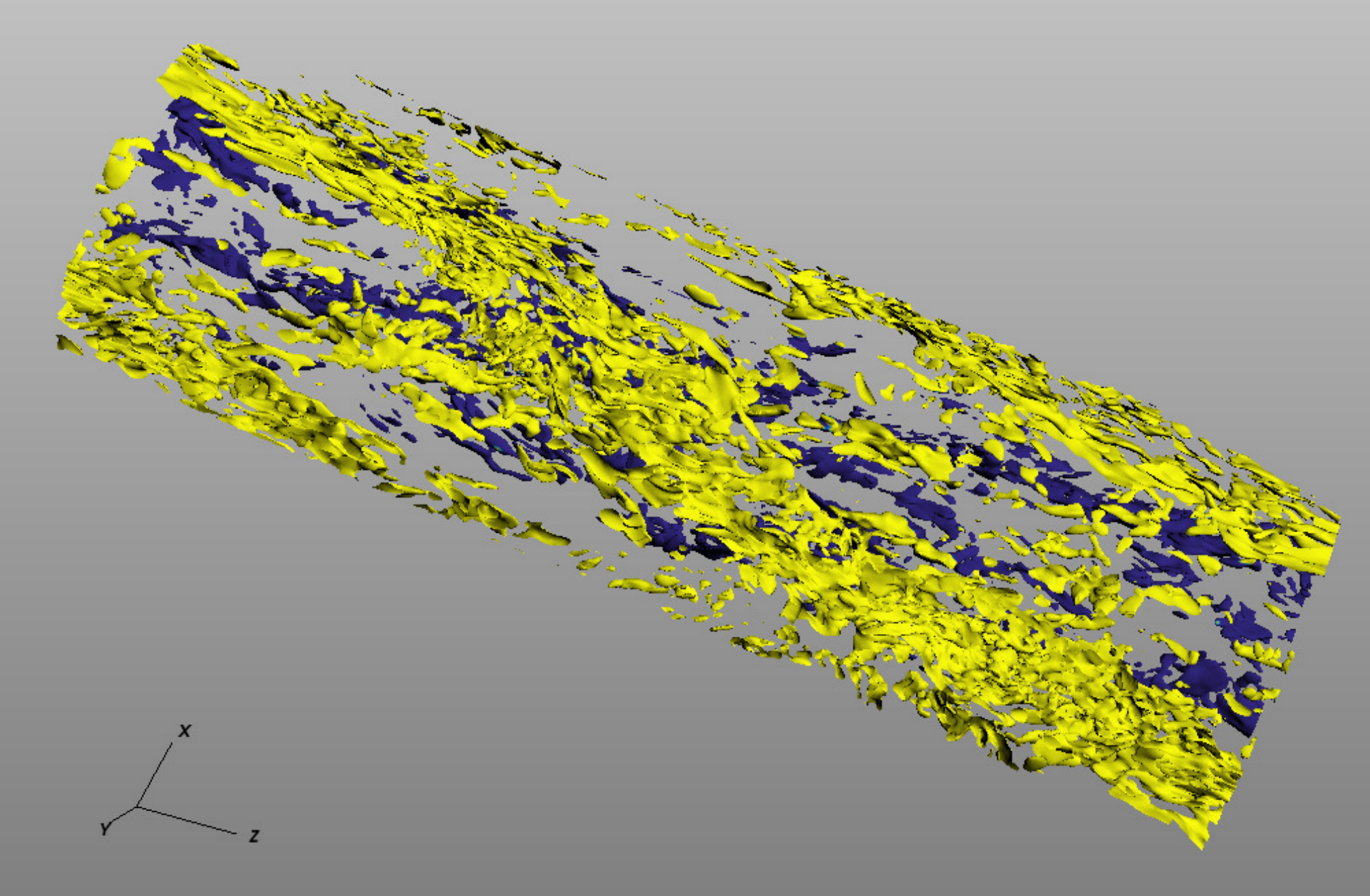}\\
  \caption{(Color online) Vortices depicted by $Q$-criterion in the region $1.2<r<1.8$ for Case C2. Blue color (dark grey) marks the vortices in region $1.2<r<r_0$ and yellow color (light grey) in region $r_0<r<1.8$, respectively. The main axial flow is from upper-left corner to lower-right corner.}\label{fig:fort}
\end{figure}

The flow structures associated with IWs are shown in Fig.~\ref{fig:fort} for Case C2, where the vortices are depicted in the region $1.2<r<1.8$ by the $Q$-criterion~\cite{Hunt_etal:1988}. The yellow (or the light grey) color marks the vortices at the region $r>r_0$ and the blue (or the dark grey) color at the region $r<r_0$, respectively. The mean axial flow is from the upper-left corner to the lower-right corner. Clearly, the vortices at $r>r_0$ form very large clusters which tilt away from the axis by a small angle. The vortices at $r<r_0$ also form some weak clusters, which tilt towards the opposite direction to those clusters at $r>r_0$. Since we employ the periodic conditions in both azimuthal and axial directions, those clusters are actually helix-like and rotate around the inner boundary of the annulus. These structures share the same nature as those in the streamwise rotating channel flow~\cite{Yangetal:JFM:2010}, where two sets of large scale clusters locate at each side of the centreline where the mean axial velocity reaches maximum, since the streamwise rotating channel can be treated as an annulus with infinity large radius. 

\section{Conclusions}\label{sec:conclusion}

In summary, we have found the inviscid IW solutions for circular pipe or annulus which constantly rotates around its axis with a moderate angular speed. Our analysis shows that even when there is an axial mean flow, the IW solutions can still exist in the bulk region provided that the mean velocity profiles satisfy certain IW conditions. Namely, the mean azimuthal velocity profile obeys the zeroth-order Bessel function of the first kind and the mean axial velocity profile takes the shape of the first-order Bessel function of the first kind, respectively. The mean velocity acts as a modulation of the background rotation.

The theory is then validated against the numerical results of rotating pipe and annulus flow. Indeed in the bulk region the mean velocity profiles exhibit the exact shapes which the theory predicts for both flows. Compared to the non rotating case, the existence of IWs induces a mean azimuthal flow and modifies the mean axial flow, so that the mean velocity profiles evolve into certain shapes and satisfy the IW condition. Flow visualisation reveals that these IWs are associated with large scale vortex clusters, which is very similar to that in streamwise rotating channel flow~\cite{Yangetal:JFM:2010}. All these findings for different domains suggest that such IWs are quite common in that rotating wall bounded flow with rotation axis being parallel to the main flow direction. And these IWs are strong enough to modify the mean velocity profiles.

\section*{Acknowledgement}

The authors appreciate the valuable comments of the anonymous referees. YY and RO are partially supported by FOM and ERC. YY and JW also acknowledge the partial support by the National Natural Science Foundation of China, Key Project No. 10532010. Computational time for the pipe DNS was given by CASPUR and for the annulus by a PRACE grant in resource CURIE. 

\appendix
\section{Helical wave functions in cylindrical coordinates}\label{app:hwf}

We denote the radial, azimuthal and axial coordinates by $(r,\theta,z)$ and the corresponding unit vectors by $(\bm{e}_r, \bm{e}_\theta, \bm{e}_z)$, respectively. In the axial direction, periodicity is used. The helical wave functions $\bm{\phi}_s$ are the eigenfunctions of the curl operator with non-penetrating condition at wall boundary, i.e.
\begin{equation}\label{eq:eigenprob}
  \left\{\begin{array}{lcl}
    \displaystyle \nabla\times\bm{\phi}_s(k,\,\bm{x}) & = & sk \bm{\phi}_s(k,\,\bm{x}),   \\[0.2cm]
    \displaystyle \bm{e}_r\cdot\bm{\phi}_s(k,\,\bm{x}) & = & 0,   \quad\quad \mbox{at\ wall boundary}.
  \end{array} \right.
\end{equation}
Here $s=\pm 1$ is the polarity index, $k$ is the wave number, and $sk$ is the eigenvalue associated with $\bm{\phi}_s(k,\,\bm{x})$, respectively. These functions possess interesting properties. Each of them is a Beltramian field with its curl parallel to itself everywhere. Thus they are steady solutions of the inviscid Euler equation. These functions have been previously applied to three dimensional incompressible flow as the ``Beltramian spectrum''~\cite{Constantin_Majda:1988}. Moreover, Yoshida and Giga~\cite{Yoshida_Giga:1990} have proved that for single-connected domain $D$, these functions form a complete basis set for any divergence-free field and satisfy the functionally orthogonal relationship
\begin{equation}\label{eq:ortho}
    \int_D \bm{\phi}_s(p)\cdot\bm{\phi}^*_t(q) dV = \delta_{st}\,\delta_{p\,q}
\end{equation}
with $\delta$ being the Kronecker delta. The helical wave functions are functionally orthogonal to each other for both different polarities and wave numbers. 

The analytical solutions of \eqref{eq:eigenprob} may be derived for some simple domains by using the construction formula~\cite{Chand_kendall:1957}
\begin{equation}\label{eq:CK}
    \bm{\phi}_s = \nabla\times(\bm{e}_z \psi) + \frac{s}{k}\nabla\times\nabla\times(\bm{e}_z \psi),
\end{equation}
where $\psi$ satisfies the Helmholtz equation
\begin{equation*}
    \nabla^2 \psi + k^2 \psi = 0.
\end{equation*}
We can readily find that~\cite{WMZ:2006}
\begin{equation*}
  \psi = J_m(k_r r) \exp[i(m\theta + k_z z)],
\end{equation*}
with $i=\sqrt{-1}$, $m=0,1,2,...$, $k_z$ is the axial wave number, and $k_r^2=k^2-k_z^2$, respectively. $J_m$ is the $m$-th order Bessel function of the first kind. The boundary condition for $\psi$ can be determined by ensuring $\bm{e}_r\cdot\bm{\phi}_s = 0$ at the wall boundary. For cylinder or circular annulus, the helical wave functions take the form
\begin{equation}\label{eq:basis}
    \bm{\phi}_s  =  \bm{\varphi}_s(r;\,k_r,m,k_z)\exp[i(m\theta + k_z z)].
\end{equation}

\end{document}